\documentclass[12pt,a4paper]{article}

\usepackage[T1]{fontenc}

\usepackage[utf8]{inputenc}

\usepackage{a4wide}
\usepackage{latexsym}
\usepackage{epsf}
\usepackage{amssymb}
\usepackage{graphicx}
\usepackage{amsmath, cite}
\usepackage{amsmath,amssymb,amsthm, slashed}
\usepackage{verbatim}
\usepackage{hyperref}
\usepackage{color}	
\usepackage{bbold}

\usepackage{tikz}
\usetikzlibrary{shapes.geometric, arrows}

\newcommand{\x}{\times}

\def\te{\tilde \epsilon}

\newcommand{\be}{\begin{equation}}
\newcommand{\ee}{\end{equation}}
\newcommand{\beq}{\begin{equation}}
\newcommand{\eeq}{\end{equation}}
\newcommand{\ba}{\begin{eqnarray}}
\newcommand{\ea}{\end{eqnarray}}

\begin{document}
\numberwithin{equation}{section}

\begin{center}

{\large {\bf Classical de Sitter solutions in three dimensions without tachyons?}}

\vspace{1.1 cm} {\large   Fotis Farakos$^a$, George Tringas$^b$, Thomas Van Riet$^a$}\\

\vspace{0.5 cm} { $a$ Instituut voor Theoretische Fysica, K.U. Leuven,\\
Celestijnenlaan 200D B-3001 Leuven, Belgium.
\\[0.2cm]
$b$ Physics Division, National Technical University of Athens,\\
15780 Zografou Campus, Athens, Greece \\}

\vspace{1cm}

{\rm \footnotesize E-mails: fotios.farakos@kuleuven.be, georgiostringas@mail.ntua.gr, thomas.vanriet@kuleuven.be} 

\vspace{1cm}
{\bf Abstract}
\end{center}

\begin{quotation}
{\small 
We continue the study of compactifications of massive IIA supergravity on G2 orientifolds and demonstrate that breaking supersymmetry with anti-D2 and anti-D6 sources leads to 3d theories for which the typical tachyons haunting classical dS solutions can be absent. However for a concrete torus example the meta-stable dS window disappears after a quantization of fluxes and charges. We discuss the prospects of more general G2 compactifications and argue that they could potentially alleviate the tachyon problem by introducing larger tadpole numbers and warped throats. However, exactly those ingredients then seem to push the vacuum towards the brink of perturbative brane-flux decay in the open string sector. This is either a remarkable illustration of the no-dS swampland conjecture or such vacua live in very difficult to control regions of parameter space. 
}
\end{quotation}
\newpage
\tableofcontents 
\newpage

\section{Introduction}
One of the most basic requirements for undertaking string phenomenological research is moduli stabilization and controlled SUSY breaking in such a way that the extra dimensions are invisible because their fluctuations require too much energy to be detectable with current technology. Hence the search for string vacua with small extra dimensions, stabilized moduli and absence of supersymmetry remains a worthy endeavor. A conservative viewpoint would furthermore require the vacuum to be a meta-stable dS space. Such view on string phenomenology is however not the only one, and one can contemplate instead brane world models or situations in which fields roll down an effective potential. Thanks to the work on the Swampland program \cite{Brennan:2017rbf, Palti:2019pca} it became clear that there is no consensus in the community on whether string theory actually achieves a landscape of flux vacua with said properties and thus no convergence on which viewpoint will turn out to be the right path for the future of string phenomenology. 

Here we continue the scrutiny of the conservative setting; can we achieve small extra dimensions, with all moduli stabilized and SUSY broken in a dS vacuum? This question has been studied, for obvious reasons, mostly for flux vacua in 4d, see \cite{Silverstein:2004id, Grana:2005jc, Denef:2007pq, Douglas:2006es, Denef:2008wq} for general reviews and \cite{Baumann:2014nda, Danielsson:2018ztv, Cicoli:2018kdo} for reviews with an emphasis on de Sitter vacua.  It is however useful to think of other dimensions as well. From a landscape viewpoint this is anyways required if one wishes to understand what the total space of vacua is. Generic vacua can have any number of compact dimensions up to 9. It should also be obvious that there are more vacua in lower dimensions since the amount of compact manifolds, ways to wrap branes and fluxes increases dramatically with every extra compact dimensions. Very intriguingly there is not a single suggestion known to achieve moduli stabilization with small compact dimensions in $d>4$. That makes $d=4$ rather special in a flux context similarly to the string gas picture \cite{Brandenberger:1988aj}. 

In this work in particular we will consider 3d vacua. In \cite{Farakos:2020phe} we have argued that compactifications of massive IIA supergravity on G2 orientifolds with fluxes can lead to full moduli stabilization in 3d (SUSY) AdS vacua that allow a tuning to arbitrary weak string coupling, large radii and a parametric separation of scales between the AdS length and the KK scale. This is very analogous to flux compactifications of massive IIA on Calabi--Yau orientifolds to four dimensions \cite{DeWolfe:2005uu,Derendinger:2004jn}.  

Such striking parametric separation of scales at weak coupling is in contradiction with some Swampland conjectures \cite{Lust:2019zwm} (see also \cite{Blumenhagen:2019vgj}). These conjectures are loosely derived from the distance conjecture \cite{Ooguri:2006in} and inspired from no-go theorems with given assumptions \cite{Gautason:2015tig}. Recently however a suggestion was made how the massive IIA vacua can nonetheless be consistent with the web of swampland conjectures in a very interesting way \cite{Buratti:2020kda}. So purely based on the conceptual ideas surrounding the Swampland it could be that such scale separation is consistent with our understanding of string theory and the question is much open right now. 
 
On the technical side however, there have always been reasons to doubt the consistency of the massive IIA vacua \cite{McOrist:2012yc, Banks:2006hg}. These worries are related to the backreaction of the O6-planes which have only been taken into account in a \emph{smeared} fashion \cite{Acharya:2006ne}. Relatedly there is also no 11-dimensional uplift of the strongly coupled region near an O6 singularity. However, partial results about the backreaction of the localized sources are known and encouraging \cite{Saracco:2012wc, Junghans:2020acz, Marchesano:2020qvg} and simple flux vacua exist for which it can be shown that smearing is harmless on the condition there is a large volume/weak coupling limit \cite{Baines:2020dmu}.

It is our hope that having an infinite family of 3d AdS vacua with scale separation at weak coupling allows an easier holographic CFT study than with 4d vacua which hopefully shines a complementary light on these issues. 

In this paper we continue our study of compactifications of massive IIA supergravity on G2 orientifolds with fluxes but turn to the question of the existence of meta-stable dS vacua which have been conjectured to be impossible completely \cite{VanRiet:2011yc, Danielsson:2018ztv, Obied:2018sgi} or impossible at sufficiently weak coupling \cite{Garg:2018reu, Ooguri:2018wrx, Andriot:2018wzk, Bedroya:2019snp}. Our setup is entirely classical in the sense that we stick to 10d supergravity at the two-derivative level with orientifold and D-brane sources.

The quest for classical dS vacua has already a history and started with the suggestive works \cite{Silverstein:2007ac, Hertzberg:2007wc} and later the more concrete proposals of \cite{Caviezel:2008tf, Flauger:2008ad, Danielsson:2009ff}. A thorough but outdated scan and overview can be found in \cite{Danielsson:2011au} whereas an update of the recent situation is described in \cite{Banlaki:2018ayh, Junghans:2018gdb, Andriot:2019wrs, Andriot:2020wpp, Andriot:2020vlg} and \cite{Cordova:2018dbb, Cordova:2019cvf, Cribiori:2019clo,Kim:2020ysx}. Most of these works focussed on 4d vacua, but some preliminary results about higher dimensions are known \cite{VanRiet:2011yc} and a complicated suggestion for a meta-stable solution in 3d was proposed in \cite{Dong:2010pm}.

In here we follow a route towards classical dS solutions akin to \cite{Dong:2010pm} (but considerably simpler) and consider general mixtures of orientifold and anti-brane sources such that the lower-dimensional would-be EFT has no (linearly realized) supersymmetry. Of course the dangers are around every corner in that case since one should worry about the anti-brane stability as well as their backreaction. In contrast, most of the works on classical dS solutions start with calibrated orientifold and D-brane sources and break SUSY spontaneously instead of explicitly which provides slightly more control at first sight. Nonetheless we will argue for a certain amount of control directly from 10 dimensions, by verifying whether the approximations made are justified. This means that all curvature and inverse length scales should be small in string units and that the string coupling is small. In our concrete example this will be the case but not parametrically in contrast with the AdS vacua constructed in an earlier paper \cite{Farakos:2020phe}. The most surprising outcome of our analysis is that by adding \emph{two} different species of anti-branes together, namely anti-D2 and anti-D6, we can achieve dS critical points with the following properties: 1) with some tuning of coefficients we can get rid of the typical tachyons present in classical dS vacua, 2) the internal manifold does not need to be negatively curved, 3) the resulting model is very simple. Unfortunately the tuning required is impossible for the simple torus examples in this paper, but there is no reason to expect that a general G2 construction with warped throats would not allow it. Interestingly we will find evidence that exactly those ingredients could trigger perturbative brane-flux decay. This can be taken as non-trivial circumstantial evidence for the no-dS conjecture although more concrete models should be constructed to verify our general findings.

\section{Mass producing 3d de Sitter?} 
We have argued in a previous paper that G2 compactifications of massive IIA supergravity with O2/O6 sources allow the stabilization of all moduli if enough fluxes are turned on  \cite{Farakos:2020phe}. The vacua are then (SUSY) AdS$_3$ at tunably weak coupling, large length scales and separation between KK and AdS scale. It is then tempting to somehow uplift these vacua to meta-stable dS by adding SUSY-breaking ingredients with positive energy. It is known that this is not a good strategy in 4d \cite{Kallosh:2006fm}\footnote{Although see \cite{Kallosh:2018nrk}.} and the same applies in 3d. Heuristically this works as follows: AdS vacua that are well suited for uplifting have the feature that the mass $m$ of the lightest (non-axionic) scalar is large in AdS units, 
that is 
\be\label{crazy}
m^2 L^2 >> 1\,,
\ee
where $L$ is the AdS length. Such vacua are at the bottom of a scalar potential that approaches zero from below while being very narrow as depicted in figure \ref{fig_potentials} below.

\begin{figure}[h]
	\centering
	\includegraphics[width=0.7\linewidth]{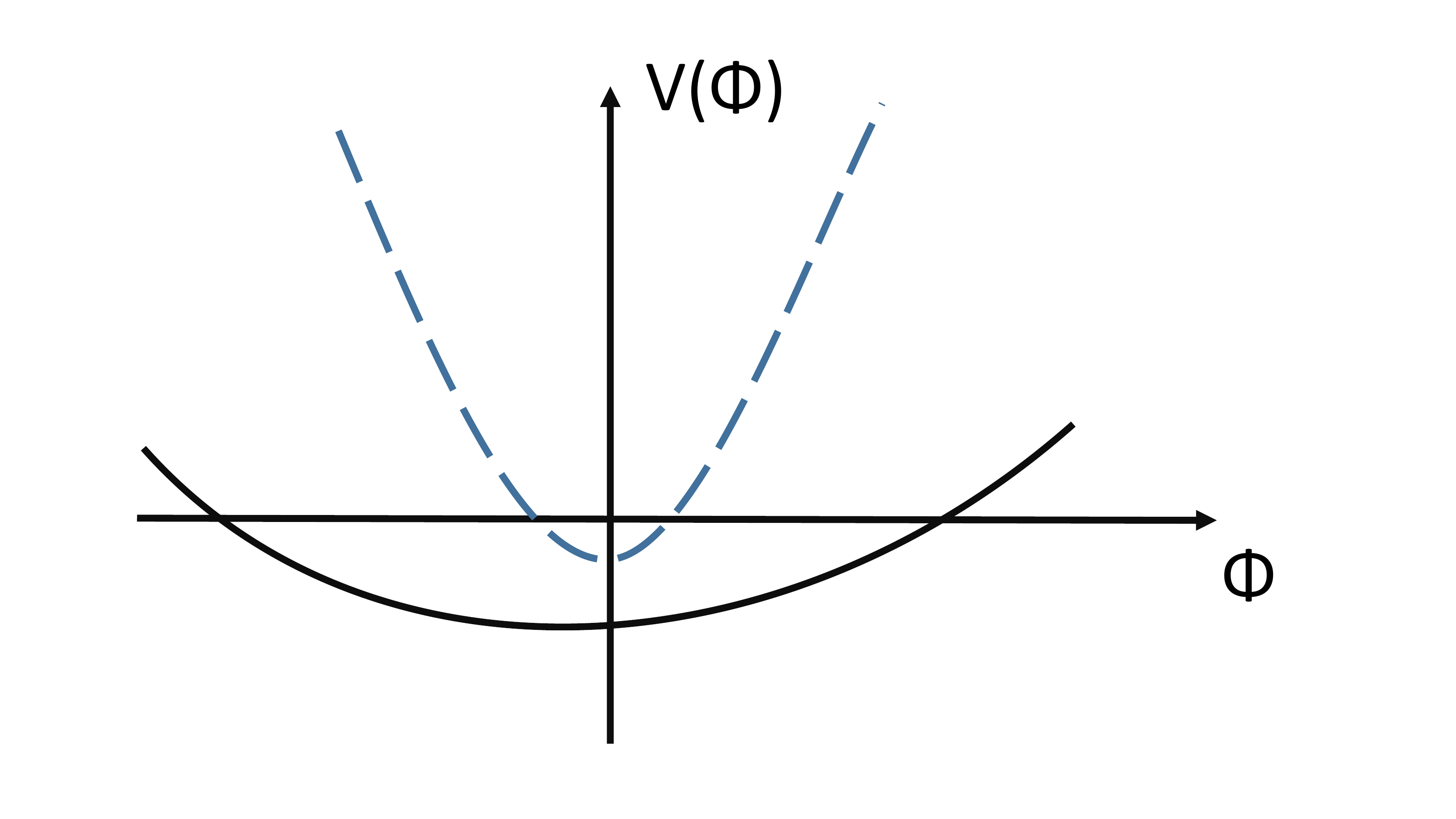}
	\caption{\small \emph{ The potential with the dashed line is better suited for uplifting than the potential with the solid line.}} \label{fig_potentials}
\end{figure}

Due to the large $L$ a small SUSY-breaking source necessarily brings one to positive energy and due to the high $m^2$ it will not destabilize the system. So models that achieve (\ref{crazy}) very well were therefore conjectured to be in the Swampland \cite{Gautason:2018gln}. This Swampland conjecture is of course inspired by the no-dS conjectures but applies to AdS vacua and should therefore be easier to prove or disprove. Despite the difficulty in finding vacua obeying (\ref{crazy}) this conjecture is furthermore inspired by the bizarre properties the dual CFT would have since $m^2L^2$ determines the dual conformal operator weight. The KKLT and LVS AdS vacua \cite{Kachru:2003aw, Balasubramanian:2005zx} are our most concrete suggestions for vacua that get close to obey (\ref{crazy}), but they cannot do so tunably.  Racetrack models are built to achieve (\ref{crazy}) if they would allow the ``Kallosh--Linde'' fine-tuning \cite{Kallosh:2004yh}, but there is not a single string theory example showing such behavior and it might also be in tension with Swampland bounds on axion decay constants \cite{Moritz:2018sui}, see however \cite{Kallosh:2019axr, Blanco-Pillado:2018xyn}. Finally it has been demonstrated that non-geometric flux backgrounds can achieve (\ref{crazy}) arbitrary well because they allow moduli-stabilized Minkowski solutions \cite{Micu:2007rd, deCarlos:2009qm}, where effectively $m^2 L^2=\infty$. But such backgrounds are far from being shown to be trustworthy because the EFTs derived from non-geometry are difficult to control, see \cite{Plauschinn:2018wbo} for a review.

Although KKLT and LVS do not obey (\ref{crazy}) tunably it is suggested they achieve it sufficiently well to allow anti-D3 uplifts to dS vacua. However that procedure would require warped throats in order to make the uplift energy tunably small, but recently it has been appreciated that demanding the throat volume to fit inside the total Calabi--Yau volume is so constraining that the tuning freedom might be lost \cite{Carta:2019rhx}.  We will come back to issues related to gluing throat regions into compact spaces later.  

This paper is about following a somewhat related strategy but in 3d and with only classical\footnote{Here `classical' refers to string theory ingredients whose leading order contributions to the energy can be captured using 10d supergravity at the two-derivative level, with inclusion of source terms.} ingredients (fluxes, branes and orientifolds).  So we turn to massive IIA supergravity with O2/O6-planes, (anti) D2/D6-branes. 
We will not attempt to uplift the AdS vacua of \cite{Farakos:2020phe}, 
but we will instead search directly for meta-stable de Sitter critical points arising from uplifting moduli-stabilized non-SUSY Minkowski critical points. 
Obviously any non-SUSY Minkowski minimum that is not of the no-scale type will arise from a fine-tuning that is almost certainly impossible after quantization of fluxes and charges, but we will use it nonetheless as a guiding principle and afterwards compute what the effects of the quantization are. For the sake of meta-stable dS$_3$ solutions quantization will turn out problematic for simple set-ups but we argue that we do not expect this to be a real issue for more involved set-ups.

We start our analysis by restricting to the universal bulk moduli, that is the dilaton and the volume.  Generically these scalars are the typical place where the classical tachyonic instability would show up  when we study Minkowski/de Sitter vacua with broken and full moduli stabilization.\footnote{And if not in that 2-scalar subsector, the tachyon is in a 3-scalar sector with the third scalar representing the overall volume of the  orientifold cycles\cite{Danielsson:2012et}.}
As we will see this situation is seemingly not the case here because these tachyons don't show up in a generic setup. A reason for why we can avoid tachyons is related to our discussion around figure \ref{fig_potentials} as the ``Minkowski limit'' of our models is not of the no-scale type, see also \cite{Junghans:2016abx}. 

The two universal bulk moduli we have here are the real scalars $x$ and $y$ 
and they are a linear combination of the dilaton and the volume modulus defined in \cite{Farakos:2020phe} and equation (\ref{redefs}) further below.
The scalar potential then reads\footnote{If we had included the curvature of the internal manifold its contribution 
would be $V_{R7} \sim - R_{7} \, e^{\frac{3y}{2}  - \frac{x}{2 \sqrt 7}}$. 
However here we work directly with a Ricci flat internal space therefore $R_{7}=0$.} 
\be
\label{rho-tau}
V = 
A \, e^{2 y} 
+ F \, e^{2y - \frac{2 x}{\sqrt{7}}} 
+ H \, e^{2y + \frac{2 x}{\sqrt{7}}} 
+ C \, e^{y - \sqrt{7} x }  
+ T \, e^{\frac{3y}{2}  - \frac{5 x}{2 \sqrt{7}}}  \, , 
\ee
where we have used the symbols $A, F, H, C, T$ to indicate the various contributions from the fluxes and the sources. These are functions of all other scalar fields specific to a compactification. 
We will give the exact origin of these terms and their form for a toroidal orbifold in the next section. 
The coefficient $A$ relates to the O2-plane tension and $T$ to the O6-plane tension and both are negative. 
The coefficients $F$, $H$ and $C$ are related to $|F_4|^2$, $|H_3|^2$ and $|F_0|^2$ respectively, and they are positive definite. 
For technical simplicity we also absorb the vacuum values of $x$ and $y$ in these coefficients such that 
\be
\langle x \rangle = 0 = \langle y \rangle \, . 
\ee
Similar to \cite{Silverstein:2007ac, Hertzberg:2007wc} we then analyze the three conditions 
\be
\label{VACXY}
V_x = 0 \ , \quad V_y = 0 \ , \quad V = \epsilon \, , 
\ee
where we use $\epsilon > 0$ to parametrize the  small vacuum energy. 
The ``Minkowski limit'' is therefore $\epsilon=0$ and we will consider it essentially as a crude estimation to test stability, 
and building on that, 
only a small uplifting will then give meta-stable de Sitter. 
Once we apply these equations to the scalar potential we get after few manipulations 
\be
\label{DSMcond}
\begin{aligned}
A & = -8 \epsilon - 2 F   \, , 
\\
H & = F  + C  + 5 \epsilon \, , 
\\
T  & = 4 \epsilon - 2 C  \, . 
\end{aligned}
\ee
Note that the consistency of these vacua simply requires $H > C$, $H > F$ , $A <0$ and $T <0$ because we are assuming  small $\epsilon$. 
The mass matrix of the $x$ and $y$ scalars has eigenvalues given by 
\be
m_{\pm}=\frac{2}{7}\Big( 9 C +2 H 
\pm\sqrt{88 C^2 +29 C H + 4 H^2}\Big) 
+ {\cal O}(\epsilon) \, , 
\ee
where we have used the vacuum conditions. 
Since we can make $\epsilon$ arbitrarily small, 
to check the positivity of the masses we only need to look at the $\epsilon$-independent parts (i.e. the Minkowski limit). 
The reader can verify that 
\be
\label{xyEigen}
m_{\pm} > 0 \ \ \to \ \ H > C \, , 
\ee
which is in complete agreement with the consistency of the de Sitter solutions. 

We thus conclude that we have at hand a classical framework for ``mass production'' of 3d de Sitter. The reason it works (at the level of the 2 universal scalars) is exactly because of our arguments surrounding figure \ref{fig_potentials}: we have assumed small $\epsilon$ (effectively put it to zero) and found a positive mass matrix in the universal directions. So the presence of such regions in the scalar potential implies there are good reasons to expect meta-stable vacua. 
  
We will see however for our specific example that careful consideration of quantization and tadpole conditions changes this naive estimate 
and makes us wonder if these vacua are in the swampland instead. 
Indeed notice first that if we did not take $\epsilon$  small then we could have instabilities 
because the leading order contribution of $\epsilon$ to the masses is 
\be
\label{MMdet}
7\, m_+ \, m_- = 4 C ( H - C) 
- 4 \epsilon \, (39 C + 4 H  ) 
+  {\cal O}(\epsilon^2)  \, . 
\ee
From the second term we can see why an instability can arise if the de Sitter vacuum is not shallow enough, 
i.e. if $\epsilon$ is not small 
enough\footnote{Notice that had we set $F=0$ then $H-C$ would be of order $\epsilon$ and so the expression \eqref{MMdet} would not be reliable. 
In this sense we are implicitly assuming that $F$ has a considerable contribution. 
However in our examples latter we explicitly check that the masses are positive in any case thus verifying our generic analysis.}. 
In fact due to flux quantization the vacuum conditions \eqref{DSMcond} put strong constraints on the possible values of $\epsilon$, 
especially once we take the tadpole cancelation conditions into account which bound the fluxes. 
Therefore having arbitrarily small $\epsilon$ should not be taken for granted. 

Note that our analysis has not used any curvature of the internal 7D manifold, which would contribute another piece to the 3d potential. The fact that the equations $\partial_x V=\partial_yV=0$ are consistent with the absence of such a term is different from the usual classical dS vacua constructions. The reason this is possible now is because of a \emph{mixture} of different brane types, i.e., O2/D2 and O6/D6, which is usually not considered. One could worry that a Ricci-flat Ansatz is not consistent with the 10d equations of motion due to backreaction of fluxes. However the self-consistency of the approach is guaranteed if a critical point is found and from a 10d viewpoint one finds that the negative tension of the smeared orientifolds exactly cancels the positive tension of the branes and fluxes inside the compact dimensions. 
Whether or not the smearing of the orientifolds is a problem depends probably on how small the coupling can be 
and how large the internal volume is \cite{Baines:2020dmu}.

\section{A toroidal example}

In the previous section from the generic form of the scalar potential 
we have argued that massive IIA compactified on a 7D Ricci flat space with O-plane and anti-D-brane sources could potentially lead to meta-stable de Sitter vacua. 
Now we attempt at finding  a specific example: we specify the ingredients, 
we derive the potential \eqref{rho-tau}, 
and we also discuss the tadpoles and flux quantization conditions. 
We set $\alpha'=1$, we utilize the orientifold and orbifold  of \cite{Farakos:2020phe}, 
and we focus directly on solutions with an isotropic seven-torus for simplicity but also as we will show, it can guarantee stability of the non-universal scalars.

\subsection{Toroidal orientifold recap}

Before venturing into the fluxes and the moduli let us review the toroidal orientifold (i.e. the singular G2) used	 in \cite{Farakos:2020phe}. 
We shall refer to this internal 7d space as $X$. 
To construct it we consider a seven-torus $\mathbb{T}^7$ with coordinates $y^m \simeq y^m+1$ and we 
orbifold over the group of $\mathbb{Z}_2$ involutions $\Gamma=\{\Theta_\alpha,\Theta_\beta,\Theta_\gamma\}$. 
These involutions are defined as 
\be
\begin{aligned}
\Theta_\alpha : (y^1, \dots, y^7 ) & \to (-y^1, -y^2, -y^3, -y^4, y^5, y^6, y^7) \, , 
\\
\Theta_\beta : (y^1, \dots, y^7 ) & \to (-y^1, -y^2, y^3, y^4, -y^5, -y^6, y^7) \, ,
\\
\Theta_\gamma : (y^1, \dots, y^7 ) & \to (-y^1, y^2, -y^3, y^4, -y^5, y^6, -y^7) \, , 
\end{aligned}
\ee
and one should also include all their combinations, 
e.g. $\Theta_{\alpha\beta} = \Theta_{\alpha}\Theta_{\beta}$, etc., 
in the group $\Gamma$. 
We do not resolve the singularities and as a result we are working with a singular G2 space for which $b_3=7$. The calibrated 3-form is given by 
\be 
\Phi = \sum_{i=1}^7 s^i(x) \Phi_i \, , 
\ee 
where the $s^i(x)$ describe the internal metric deformations and relate to the vielbeins as $e^1 = (s^1 s^6 s^7)^{1/2} (\prod_i s^i)^{-1/6}  dy^1$, etc. 
The basis of harmonic 3-forms is given by 
\be
\Phi_i = \{ dy^{127}, - dy^{347}, - dy^{567},  dy^{136}, - dy^{235},  dy^{145},  dy^{246} \} \, , 
\ee 
where $dy^{127} = dy^1 \wedge dy^2 \wedge dy^7$, etc. 
We also make use of a basis of harmonic four-forms $\Psi_i$ which are defined by 
\be 
\int_7 \Phi_i \wedge \Psi_j =\delta_{ij} \, . 
\ee  
Finally the volume is given by 
\be 
\text{vol}(X) = \left( \prod_{i=1}^7 s^i \right)^{1/3} = \frac17 \int \Phi \wedge \star \Phi \, , 
\ee 
where we use $\int_{\mathbb{T}^7} dy^1 \wedge \dots \wedge dy^7 =1$ in the covering space.

Now we turn to the space-filling O2-planes. 
We consider the target space part of the O2 action, denoted $\sigma$, as the following $\mathbb{Z}_2$ involution 
\be
\sigma : (y^1, \dots, y^7 )  \to (-y^1, -y^2, -y^3, -y^4, -y^5, -y^6, -y^7) \, . 
\ee  
The $\sigma$ has $2^7$ fixed points in the torus covering space. 
The G2 calibration is odd under the O2 involution 
\be
\sigma : \Phi \to - \Phi \, , 
\ee 
and also note that the $\Gamma$ and the $\sigma$ commute. 
Moreover, we ask that the orbifold image of an O2 is again some physical object 
and as a result we find in total 7 different directions for O6-planes 
\begin{align}
\begin{pmatrix} 
&{\rm O}6_{\alpha}:\quad & \times & \times & \times & \times & - & - & -  \\
&{\rm O}6_{\beta} :\quad &\times & \times & - & - & \times & \times & -  \\
&{\rm O}6_{\gamma}:\quad &\times & - & \times & - & \times & -& \times  \\
&{\rm O}6_{\alpha\beta}:\quad & - & - & \x & \x & \x & \x & -  \\ 
&{\rm O}6_{\beta\gamma} :\quad & - & \x & \x & - & - & \x & \x  \\ 
&{\rm O}6_{\gamma\alpha} :\quad &- & \x & - & \x & \x & - & \x  \\ 
&{\rm O}6_{\alpha\beta\gamma}:\quad &\x & - & - & \x & - & \x & \x 
\end{pmatrix} \, , 
\end{align}
that are defined by the respective combination of $\sigma$ and $\Gamma$. 
For example one can see that the target space part of the involution 
induced by ${\rm O}6_{\alpha}$ is $\sigma \Theta_{\alpha}$.  
The target space part of each O6 involution has $2^3$ fixed points. 
From here we see that the total O6 source 3-form $J_3$ appearing in the $F_2$ Bianchi identity will be given by 
\be
J_3 = \sum_i \Phi_i \, . 
\ee

These involutions reduce greatly the amount of remaining supersymmetry in 3 dimensions. 
Firstly the orbifold $\Gamma$ reduces the ``N=16'' (in 3d spinors) 
of the massive IIA to N=2 in three dimension, 
and then the O2 further reduces the surviving supersymmetry to N=1. 
As a result the formulation of this theory becomes purely real. 
Indeed, 
the only closed string scalars are the $s^i$ that come from the metric, 
and the dilaton $\phi$. 
The second Betti number $b_2$ vanishes so there are also no axions coming from $C_3$. 
Once we add anti-branes and do not include the degrees of freedom that allow anti-branes to decay against fluxes the 3d theory has no linear supersymmetry. Of course in the open string sector there will be scalars corresponding to anti-D2 and anti-D6 positions - such scalars will have compact field ranges and tend to be stabilized in regions of high warping, if any.

\subsection{Tadpoles, flux quantization and potential}

Let us first look at the Bianchi identities which we require to satisfy without using D2- or D6-branes. 
For the O2 Bianchi identity we have 
\be
0= \int_7 H_3 \wedge F_4 - (2 \pi)^5 \left(N_{\overline{\rm D2}} + 16 \right)  \, ,  
\ee
where we have used $N_{\rm O2} = 2^7$. 
For the O6-planes we have $N_{\rm O6} = 2^3$ per 3-cycle and we allow the same number ($N_{\overline{\rm D6}}^{(i)}$) of anti-D6-branes per i'th 3-cycle:
\be
N_{\overline{\rm D6}}^{(i)} \equiv N_{\overline{\rm D6}} \ , \quad {\text{total number of}} \ N_{\overline{\rm D6}} = 7 \times N_{\overline{\rm D6}} \, .  
\ee 
Therefore for the O6 Bianchi we have for each 3-cycle 
\be
0= \underset{3-cycle}{\int} H_3 \wedge F_0 
- 2 \pi \left( N_{\overline{\rm D6}} 
+ 16 \right) 
\underset{3-cycle}{\int} J_3 \, ,  
\ee
where $J_3$ is the unit-normalized 3-form source: $J_3 = \sum_i  \Phi_i$. 
Fluxes, consistent with a isotropic seven-torus are the following:
\be
H_3 =  (2 \pi)^2 K \sum_i \Phi_i \ , \quad F_0 = (2 \pi)^{-1} M \ , \quad  F_4 = (2 \pi)^3 G \sum_i \Psi_i \, , 
\ee
where $K,M,G \in \mathbb{Z}$. 
The $F_2$ tadpole condition gives the same result for each 3-cycle:
\be
\label{O6-TAD}
16  = K M - N_{\overline{\rm D6}} \, . 
\ee
Whereas the $F_6$ tadpole implies 
\be
\label{O2-TAD}
16 = 7 K G - N_{\overline{\rm D2}} \, . 
\ee

We now spell out the various contributions to the 3d scalar potential.  
The contribution from the fluxes to the 10d action in Einstein frame (with $F_0=m$, the Romans mass) is 
\be
S_{\rm kin. \, flux} = \int_{10} \sqrt{-g_{10}} \left( R_{10} - \frac12 e^{-\phi} |H_3|^2 
- \frac12 e^{\phi/2}  |F_4|^2   
-  \frac12 e^{5 \phi/2} m^2 \right) \, , 
\ee
where we kept the ten-dimensional Ricci scalar to keep track of normalizations, 
the contributions from the sources are  (ignoring open string moduli)\footnote{Note that our p-brane actions are multiplied with an overall factor $(2 \pi)^7$ because the total action we are using is $(2 \pi)^7 \times (S_{IIA} + S_{sources})$. } 
\begin{align}
& S_{\rm p=2} = - (2 \pi)^7 \left( \mu_{\rm O2} + \mu_{\overline{\rm D2}}  \right) \, e^{-\phi/4} \int_{3} \sqrt{-g_3} \, , \\
& S_{\rm p=6} = - (2 \pi)^7 \left( \mu_{\rm O6} + \mu_{\overline{\rm D6}}  \right) \, e^{3 \phi/4} 
\sum_{i=1}^7  \int_{M_i} \sqrt{-g_7}   
\, , 
\end{align} 
where the $M_i$ denote the 7D worldvolumes of the anti-D6 branes and where 
\begin{align}
& \mu_{\overline{\rm D2}} =N_{\overline{\rm D2}} \, (2\pi)^{-2} \ , \quad 
\mu_{\rm O2} = - (2)^{-3} N_{\rm O2} \, (2\pi)^{-2} = - 16 \, (2\pi)^{-2} \, , \nonumber \\
& \mu_{\overline{\rm D6}} = N_{\overline{\rm D6}} \, (2\pi)^{-6} \ , \quad 
\mu_{\rm O6} = - 2 N_{\rm O6} \, (2\pi)^{-6} =  - 16 \, (2\pi)^{-6} \, . 
\end{align}
When one considers spaces with warped regions then  the anti-D2-brane tension can redshift  and we postpone the discussion of this until later.

Starting now from 10d Einstein frame we finally perform a direct dimensional reduction: 
\be
ds_{10}^2 = e^{2 \alpha v} ds_3^2 + e^{2 \beta v} \widetilde{ds}_7^2 \, , 
\ee
where $\alpha^2=7/16$, $\alpha=-7\beta$ and $\widetilde{ds}^2_7$ is the metric on a unit-volume G2 space. Following \cite{Farakos:2020phe}, 
we perform a rescaling of the 3d metric $g_{\mu\nu} \to \frac14 g_{\mu\nu}$ to match to more conventional units for 3D supergravity theories.  We extract the volume from the metric deformation moduli by setting 
\be
s^i = {\text{vol}}(X)^{3/7} \tilde s^i \ , \quad {\text{vol}}(X) = e^{7 \beta v} \, , 
\ee
where the unit-volume deformations satisfy 
\be
\prod_{i=1}^7 \tilde s^i =1 \ \to \  \tilde s^7 = \prod_{a=1}^6 (\tilde s^a)^{-1} \, . 
\ee 
Notice for example that because of these definitions 
(taking into account that $\tilde \star \Psi_i =  (\tilde s^i)^{2} \Phi_i$, where the $i$ does not sum) we have 
\be
\begin{aligned}
|F_4|^2 & =  F_4 \wedge \tilde \star F_4 = (2 \pi)^6 G^2 \sum_{i,j} \Psi_i \wedge \tilde \star \Psi_j 
= (2 \pi)^6 G^2 \sum_{i=1}^7 (\tilde s^i)^2 
\\
& = (2 \pi)^6 G^2 \left( \sum_{a=1}^6 (\tilde s^a)^2 + \prod_{a=1}^6 (\tilde s^a)^{-2} \right) \,, 
\end{aligned}
\ee
and similarly for $|H_3|^2$ and the other terms. 
We combine the volume and the dilaton in the following useful combinations 
\be
\label{redefs}
\frac{x}{\sqrt7} = - \frac{3\phi}{8} + \frac{\beta}{2}v\,,\qquad 2y = -21\beta v - \frac{1}{4}\phi \,. 
\ee 
These are the $x$ and $y$ universal moduli we discussed in the previous section. 
Once we include all these ingredients, 
the 3d scalar potential becomes 
\be\label{potential1}
V = 
A e^{2 y} 
+ F(\tilde s^i) e^{2y - \frac{2 x}{\sqrt{7}}} 
+ H(\tilde s^i) e^{2y + \frac{2 x}{\sqrt{7}}} 
+ C e^{y - \sqrt{7} x }  
+ T (\tilde s^i) e^{\frac{3y}{2}  - \frac{5 x}{2 \sqrt{7}}}  \, , 
\ee 
with the coefficients given by 
\be
\begin{aligned}
A  & = \frac{(2 \pi)^5}{8} \left( 2 N_{\overline{\rm D2}}  - 7 K G  \right) e^{2 y_0} \,, 
\\
F & = \frac{(2 \pi)^6 G^2}{16} \left( \sum_{a=1}^6 (\tilde s^a)^2 + \prod_{a=1}^6 (\tilde s^a)^{-2}  \right)  e^{2y_0 - \frac{2 x_0}{\sqrt{7}}} \,, 
\\
H & = \frac{(2 \pi)^4 K^2}{16} \left( \sum_{a=1}^6 (\tilde s^a)^{-2} + \prod_{a=1}^6 (\tilde s^a)^{2}  \right)  e^{2y_0 + \frac{2 x_0}{\sqrt{7}}} \,, 
\\
C &= \frac{M^2}{16 (2 \pi)^2} e^{y_0 - \sqrt{7} x_0 } \, , 
\\
T & = \frac{2 \pi}8 \left(2 N_{\overline{\rm D6}} - K M  \right) \left( \sum_{a=1}^6 \frac{1}{\tilde s^a} + \prod_{a=1}^6 \tilde s^a \right) 
e^{\frac{3y_0}{2}  - \frac{5 x_0}{2 \sqrt{7}}} \,, 
\end{aligned}
\ee
where $y_0$ and $x_0$ have been inserted in the same places where $x$ and $y$ appear such that in this way the vacuum is always at 
$x = 0 = y$, 
by appropriately defining $x_0$ and $y_0$. 
For completeness let us note that the kinetic terms read 
\begin{equation}
e^{-1}\mathcal{L}_{\text{kin}} = \tfrac{1}{2}R_3 -\tfrac{1}{4}(\partial x)^2  -\tfrac{1}{4}(\partial y)^2  
-\tfrac{1}{4} \int_7 \Phi_i\wedge\tilde\star\Phi_j \partial \tilde{s}^i\partial \tilde{s}^j \,, 
\end{equation}
where $\tilde \star \Phi_i = (\tilde s^i)^{-2} \Psi_i$  (no summation over $i$ implied).  The conditions \eqref{VACXY} and \eqref{DSMcond} that we studied in the previous section should now be enforced.

\subsection{Moduli stabilization}

In this part we turn to the stabilization of the 8 universal real scalar moduli ($x,y,\tilde s^a$) of toroidal orbifold compactifications. 
The very first observation we can make right away from the 
form of the scalar potential \eqref{potential1} is that all the unit-volume toroidal moduli $\tilde s^a$ are always stabilized at 
\be
\langle \tilde s^a \rangle = 1 = \langle \tilde s^7 \rangle \,, 
\ee
giving a totally isotropic torus as we anticipated. 
In this way we always have 
\be
\frac{\partial V}{\partial \tilde s^a} \Big{|}_{\tilde s^a = 1}  = 0 \, . 
\ee
We assume these vevs for the $\tilde s^a$ in what follows. 
On this background we can then notice that 
\be
\frac{\partial^2 V}{\partial x \partial \tilde s^a} = 0 = \frac{\partial^2 V}{\partial y \partial \tilde s^a} \, , 
\ee
which means there is no mass mixing between the $x,y$ moduli and the $\tilde s^a$. 
As a result we can directly evaluate the eigenvalues for the mass matrix of the moduli $\tilde s^a$ independently. 
Using the vacuum conditions (\ref{VACXY}) one can express the second derivatives of the moduli on the scalar potential 
in terms of the functions $H$ and $C$. 
First we find that the derivatives of the functions $F(\tilde s^a), H(\tilde s^a)$ and $T(\tilde s^a)$ at the vacuum are proportional to themselves,  
i.e. 
\be 
F_{ab}= \frac{4}{7} F \left( 1 + \delta_{ab} \right)   \ , \quad 
H_{ab}= \frac{4}{7} H \left( 1 + \delta_{ab} \right)    \ , \quad 
T_{ab}= \frac{1}{7} T \left( 1 + \delta_{ab} \right)  \, , 
\ee 
where $\delta_{ab}$ is the Kronecker delta. 
As a result we can evaluate $V_{ab}$ (for $a,b=1, \dots, 6$) on the vacuum, 
and  find 
\be
\label{sEigen}
{\text{Eigenvalues}}[ V_{ab} ] = \frac17 \left( 4 F + 2 H + 4 \epsilon + 2 (H-C) \right)  \times \left\{ 7,1,1,1,1,1 \right\} \, . 
\ee
Since we already require $H>C$, 
the stability of the $\tilde s^a$ is granted on any background of the type we study here. 
In other words the tachyon that generically plagues de Sitter string vacua is making its appearance here in the dilaton-volume sector 
(i.e. $x$ and $y$) and not in the internal space deformations. 
A possible underlying reason for the positive masses of the $\tilde s^a$ scalars is that $\langle P_a \rangle = 0$ (where $P$ is the superpotential) 
so they are in some sense stabilized in their supersymmetric positions\footnote{To construct the total superpotential including the SUSY-breaking sectors 
one would have to use 3d nilpotent superfields, see e.g. \cite{Buchbinder:2017qls}.}.

Now let us focus on the explicit $x$ and $y$ stabilization because we will see it gives few more consistency conditions than our generic discussion, 
especially due to flux quantization. 
For a de Sitter vacuum we ask as before that 
\be
\label{FORDS}
V \Big{|}_{x=0=y} = \epsilon  = \frac14  \left( 2 C + T \right) \, . 
\ee 
The condition \eqref{FORDS} once combined with the scalar potential readily gives 
\be
\label{y0DS}
e^{y_0/2} = \frac{ e^{- \frac{9 x_0}{2 \sqrt{7}} } M (1 - \te)}{7 (2 \pi)^3 K \left(1 - 2 w \right)}  \, , 
\ee
where we have introduced $w, \tilde{\epsilon}$ defined through 
\be
\label{TildeE}
N_{\overline{\rm D6}} = w M K  \ , \quad \te = 32 (2 \pi)^2 M^{-2}  e^{\sqrt{7} x_0 - y_0} \, \epsilon \,. 
\ee
Then for the consistency of the solution \eqref{y0DS} we find that 
\be
w < 1/2 \, . 
\ee
The  condition $\partial V / \partial x = 0$ can be simplified 
to give
\be
e^{4x_0/\sqrt{7}} = - \frac{2 (2 \pi)^2 G^2 (1 - \te)^2}{K^2 
\left(    
-2 \te^2 
+4  (3+ 14 (w-1 )w )
+ \te (39 + 140 (w-1)w )
\right) 
}  \, , 
\ee 
where again we have to check the self-consistency of this solution. 
Since $\te$ is  small,  we only need to satisfy 
\be 
\label{wBounds}
3+ 14 (w-1 )w < 0 
\  \to \  
0.311... < w < 0.688... \, , 
\ee 
with $w$ a rational number. 
Here we can see the crucial contribution from the anti-D6-branes: 
if they are set to vanish then the solution becomes inconsistent. 
Since we already have an upper bound on $w$ we find it more convenient to readily set 
\be
N_{\overline{\rm D6}} = MK / 3 \ ,  \quad  w=1/3  \, , 
\ee
which gives 
\be
\label{x0WC}
e^{4x_0/\sqrt{7}} = \frac{2 (2 \pi)^2  G^2 (1 - \te)^2}{ 
K^2 \left(    
\frac49  
- \frac{71}9  \te  
+ 2 \te^2 
\right) 
}   \, . 
\ee
From \eqref{x0WC} we see that the solution is consistent indeed as long as $\te \ll 1$. 
Note that by setting $w=1/3$ we are assuming that $MK$ is an integer multiple of $3$. 
If we chose different values of $w$ it would still have to be a rational number, 
and thus we would have to assume $MK$ to be an integer times a bigger number than $3$ which would most probably 
be in tension with the tadpole conditions if we stay on the toroidal setup.

Finally, 
inserting the above conditions into the equation for $\partial V / \partial y = 0$ 
gives an equation that we should solve for $N_{\overline{\rm D2}}$ because we have already fixed $y_0$, 
which gives  
\be
\label{ND2}
N_{\overline{\rm D2}} = 7 KG \left( \frac12 - \frac{1}{6 \sqrt 2} 
\frac{4 - 43 \te + 18 \te^2}{
\sqrt{4 - 71  \te + 18 \te^2} 
(1 - \te) 
}  
\right) \,  . 
\ee
Equation \eqref{ND2} has a strong impact. 
Since $N_{\overline{\rm D2}}$ is integer and the values of $K$ and $G$ are bounded to be rather small due to the 
tadpole conditions it is essentially impossible to satisfy \eqref{ND2} and have $\te \ll 1$. 
To see this difficulty let us set $\te = 0$ which is the Minkowski limit, 
and we get 
\be
N_{\overline{\rm D2}} \Big{|}_{\rm Mink.} = 7 KG \left( \frac12 - \frac{1}{3 \sqrt 2} 
\right) \, , 
\ee
 which can never be satisfied for any choice of integers due to the $\sqrt 2$. 
Note that one could choose a different value for $w$ instead of $1/3$, 
but still within the bounds \eqref{wBounds}, 
such that no $\sqrt{2}$ appears in \eqref{ND2}. 
In such case one is faced with the problem that the flux quanta would acquire very large values which is in tension with our toroidal tadpole conditions. 
To summarize, 
here we see that if we had ignored tadpole cancelation or flux quantization we could not see the 
inconsistency of this solution.

Before turning to extensions let us see if the $x$ and $y$ masses and the $\tilde s^a$ masses are positive. 
From \eqref{xyEigen} and \eqref{sEigen} we know that all these positivity condition boil down to 
\be
H > C \, . 
\ee
For the positivity of the $x$-$y$ masses we also need $\te \ll 1$ as we have explained. 
One can directly verify that $H>C$ by using \eqref{y0DS} as long as $1 > 2 w > 0$ (which is satisfied for $w = 1/3$). 
Thus we see that our solution guarantees the positivity of the masses.

\subsection{Beyond the toroidal orbifold}

The toroidal orbifold example could only get us this far. 
However we have learned a few things from this simple example. 
For example we have seen that the tachyon lies in the dilaton-volume sector and not in the $\tilde s^a$ sector, 
we have seen that we need two sources of supersymmetry breaking in order to even have a chance of achieving moduli stabilization in de Sitter, 
and of course we have seen that a careful consideration of the tadpole conditions reveals possible inconsistencies.

Let us then discuss how we could go beyond a toroidal orbifold. 
To address the difficulty of consistently solving an equation like \eqref{ND2} one would have to study more general spaces. 
Either spaces that allow a much larger range for the flux quanta or spaces that induce warping to the anti-D2-brane. 
Indeed, 
if we had included anti-D2 warping, 
the scalar potential would only change in the $A$ term which would become 
\be
A = \frac{(2 \pi)^5}{8} \left( 2 \alpha N_{\overline{\rm D2}}  - 7 K G  \right) e^{2 y_0} \, , 
\ee
where $\alpha$ is the effect of the warping. 
Then our calculations would follow through exactly in the same way, 
but instead of \eqref{ND2} we would have 
\be
\label{ND2-warped}
\alpha N_{\overline{\rm D2}} = 7 KG \left( \frac12 - \frac{1}{6 \sqrt 2} 
\frac{4 - 43 \te + 18 \te^2}{
\sqrt{4 - 71  \te + 18 \te^2} 
(1 - \te) 
}  
\right) \sim \frac{7}{4} KG \,  , 
\ee
which means we could easily solve \eqref{ND2-warped} by assuming an appropriate value for $\alpha$ 
as long as $N_{\overline{\rm D2}} > 7 KG/4$ (always for small $\te$). 
This small sample calculation we did here of course does not guarantee that such procedure will work but rather it points to the possible 
extensions that may lead to a classically stable 3d de Sitter.

Let us now elaborate on a specific setup for the fluxes. 
If we assume that somehow we introduce warping then one can have for example 
\be
\label{EXAMPLE}
G=4 \, , \ K=1 \, , \ M=24 \, , \ N_{\overline{\rm D6}} = 8 \, , 
\ee
and
\be
(24)^2 \, \te =  10^{-4} \ , \quad \alpha N_{\overline{\rm D2}} = 7.40034  \ , \quad N_{\overline{\rm D2}} = 12   \, . 
\ee 
The vacuum energy is now of order $10^{-24}$ in string units whereas the $x,y,\tilde s^a$ moduli masses are all positive and are of order $10^{-17}$. 
In particular we have for the $x-y$ mass matrix 
\be
m^2_{x-y} = \left(\begin{array}{cc} 1.468.. \times 10^{-16} & 4.085.. \times 10^{-17} 
\\ 
4.085.. \times 10^{-17}  & 1.2.. \times 10^{-17} \end{array}\right) \, , 
\ee
which gives eigenvalues $1.58 \times 10^{-16}$ and $5.95 \times 10^{-19}$, 
whereas the matrix of the $\tilde s^a$ masses is totally independent 
(i.e. $\langle V_{,x,\tilde s^a} \rangle = 0 = \langle V_{,y,\tilde s^a} \rangle$) 
and has eigenvalues of order $\langle V_{,\tilde s^a, \tilde s^b} \rangle \sim 10^{-17} >0$ in agreement with our general discussion. 
For this example the string coupling and volume reads 
\be
g_s = e^\phi = 0.112... \ , \quad {\text{vol}}(X) = 1.2669 \times 10^7 \, , 
\ee
which potentially makes the supergravity approximation reliable. 
In fact taking into account the very strict constraints we have here 
it makes it hard to find other combination of fluxes in the hope to reduce $g_s$ even more to weak coupling. 
This is of course the Dine--Seiberg problem that says we cannot expect to be far from a strong coupling, which seems specific to dS, not AdS \cite{Ooguri:2018wrx, Junghans:2018gdb, Banlaki:2018ayh}. 
In any case, 
one should keep in mind that we are doing a sample computation and one should look into more suitable 
internal spaces to find realistic and trustable solutions. 

Finally, we face 
a possible problem that we elaborate on in the next section: 
generically it seems that we have 
\be\label{openstringstability}
\frac{N_{\rm anti-brane} }{{\text{flux \ quanta}}} > 1  \, , 
\ee
except for the comparison of the anti-D6 with the Romans quanta which gives 
\be
\frac{N_{\overline{\rm D6}\, \text{per 3-cycle}}}{M} = \frac13 \, . 
\ee
In the next section we explain that this may be another source of instabilities implying  the need to for internal spaces that allow larger numbers for the flux quanta.

\section{Open string instabilities?}

Anti-branes break supersymmetry because they are ``anti'' with respect to the background fluxes or orientifolds. With respect to the orientifolds they carry the same charge but opposite tension and preserve different supercharges. They are repelled from the orientifolds both gravitationally and electromagnetically and need to find a position in between the orientifolds where forces cancel out. We have not checked the details of that, but assume such positions exist due to the compactness of the internal manifold. Even if the anti-branes find such stable positions, there could still be perturbative instabilities lurking around the corner. For instance anti-branes can annihilate against surrounding fluxes \cite{Kachru:2002gs}. This is possible when fluxes induce opposite brane charges via the transgression terms in type II supergravity theories 
\be
d F_q =  H_3 \wedge F_{q-2} + Q \delta\,,
\ee
where the $\delta$ denotes a ($q+1$)-form distribution describing the localized magnetic charge of (anti-) D($8-q$) branes. When that form does not have the same orientation as $H_3 \wedge F_{q-2}$ parts of that background flux can lower their flux quanta together with $Q$ as to preserve the charge. Heuristically one assumes that some D($8-q$) branes materialized out of the flux cloud and annihilated with the actual anti-D($8-q$) branes. One could think that this process is always non-perturbative because it requires the nucleation of branes out of fluxes. But this picture is too heuristic and a more detailed approach, first pioneered by Kachru, Pearson and Verlinde (KPV) in \cite{Kachru:2002gs}, shows that this process can even be perturbative. The specific mechanism relies on brane polarization \emph{aka} the Myers effect \emph{aka} the dielectric effect \cite{Myers:1999ps}. We will not go in any details about this for the case at hand but instead draw some basic lessons from known backgrounds with brane-flux instabilities and then comment on the case at hand.

Let us start with the most well studied example of anti-D3 branes at the bottom of the Klebanov--Strassler (KS) throat. This was the situation described by KPV \cite{Kachru:2002gs}. The bottom of this throat is an $S^3$ with radius squared given by $R^2=b_0^2 g_sM$, where $b_0$ is a numerical factor close to $1$ and $M$ is the RR 3-form flux quantum piercing the $S^3$. KPV found that 
\be\label{anti-D3}
\frac{
N_{\overline{\rm D3}}
}{M}<0.08\ldots , 
\ee
otherwise the anti-branes would decay perturbatively leaving $M-N_{\overline{\rm D3}}$ SUSY D3-branes behind and one less unit of NSNS flux. Other simple models of anti-branes down throats come with similar bounds. For instance anti-M2 branes down the ``M-theory CGLP throat'' have \cite{Klebanov:2010qs}
\be\label{anti-M2}
\frac{
N_{\overline{\rm M2}}
}{M}<0.05\ldots , 
\ee
where now $M$ is an $F_4$ flux quantum. Or anti-D6 branes in environments with Romans mass require \cite{Blaback:2019ucp}
\be\label{anti-D6}
\frac{
N_{\overline{\rm D6}}
}{M}<0.5\ldots , 
\ee
where now $M$ is the Romans mass quantum. Note that our torus model has $N_{\overline{\rm D6}}/M= 1/3$ and so satisfies the inequality but not parametrically.

Note that all these inequalities were derived using brane probe actions in regimes where it is unclear they should be trusted. Not only due to possible strong coupling effects but also the classical backreaction of the anti-branes (see \cite{Bena:2009xk} for some pioneering work) could be a worry and potentially enhance the instabilities \cite{Danielsson:2014yga}. However non-trivial evidence in favor of the probe action results came from the complementary blackfold treatments carried out in \cite{Armas:2018rsy, Armas:2019asf, Nguyen:2019syc} as well as from arguments pointing to the absence of dangerous singularities due to backreaction \cite{Michel:2014lva, Cohen-Maldonado:2015ssa, Cohen-Maldonado:2016cjh, Blaback:2019ucp}. It is however quite likely that the actual bounds are a bit more strict than the probe results. 

From the inequalities \eqref{anti-D3}, \eqref{anti-M2} and \eqref{anti-D6} one could be tempted to think that in general perturbative brane flux annihilation for anti-Dk branes is prevented if 
\be\label{anti-Dk}
\frac{
N_{\overline{\rm Dk}}
}{N_{\text{flux}}}< \lambda \, , 
\ee
where $N_{\text{flux}}$ is some NSNS or RR flux integer\footnote{The flux quanta should either coming from fluxes piercing the cycle that harbors the anti-brane or is ``Poincar\'e dual'' to it.}, $N_{\overline{\rm Dk}}$ is the number of anti-Dk branes and $\lambda$ some specific number. We believe that (\ref{anti-Dk}) is morally correct but there is no general formula for $\lambda$. It will be highly dependent on the specific example, the details of the manifold, the fluxes, etc. 
For instance anti-branes as defined earlier can be completely stable against brane-flux decay in AdS backgrounds \cite{Apruzzi:2013yva, Junghans:2014wda, Gautason:2015ola}. In general we expect $\lambda$ to depend non-trivially on the values of the stabilized moduli and this is why a case by case analysis is necessary. So it is impossible for us to discuss the general constraints from brane-flux decay until we have a specific model in which the closed string tachyons are absent after quantization of fluxes is taken into account. As we argue in this paper our torus example at least has shown there is no fundamental reason to expect tachyons to be present in more involved models, unlike the situation with classical dS vacua without anti-branes \cite{Danielsson:2012et, Junghans:2016uvg}. In what follows we will stick to explaining why we expect a dependence of $\lambda$ on stabilized moduli. The only exception being the case of anti-D6 branes where (\ref{anti-D6}) seems independent of details. Our strategy will be to use the most well-studied example of anti-D3 branes and demonstrate how moduli dependences creep into $\lambda$. 

We start with demonstrating $g_s$ dependences. For instance in \cite{Gautason:2016cyp} anti-D3 decay in the ``S-dual'' KS throat\footnote{This is the weakly coupled supergravity solution obtained by S-dualizing the KS solution and then dialing to small coupling.} was studied. This decay is against RR fluxes instead of NSNS fluxes and find 
\be
\frac{g_s N_{\overline{{\rm D3}}}}{K}<0.08\ldots , 
\ee
with $K$ the NSNS flux quantum. Furthermore there will be a dependence on the size of the cycle that harbors the anti-brane. In the examples of warped throats this is not obvious since the cycles at the tip are localized cycles and do not change their volume when the overall volume of the compactification manifold adjusts. So in models without anti-branes living near deformed conifolds there is a dependence. Using the KPV computation we can compute this by keeping the cycle size $L$ arbitrary.  From the brane-polarization potential in \cite{Kachru:2002gs} one then finds that
\be
\pi\left(\frac{N_{\overline{D3}}}{M}\right)_{\text{max}} =  \text{arccot}\left(\frac{L^2}{g_sM}\right) - \frac{1}{2}\sin\left(2 \text{arccot}\left(\frac{L^2}{g_sM}\right)\right)\,.
\ee
So when $L^2 = b_0^2 g_sM$ we will find (\ref{anti-D3}), but if the cycle size is set by different effects one can easily infer that large volume and small coupling enhances the decay.  This simple observation shows that using anti-branes that do not live at the bottom of warped throats generated by deformed conifolds, is risky. In fact the above formula even suggests the problems worsen at weak coupling and large volume. Hence, similar to our discussion about closed string stability we are led to inserting anti-branes in warped throats from desingularized conifolds since then these complicated dependences might be washed away. Most studies on brane-flux decay assume there are no ``compactification effects'' on the brane-flux decay when the throats are inserted in compact spaces. Although this is still somewhat unclear and preliminary results can be found in \cite{Frey:2003dm, Brown:2009yb}. Interestingly other compactification effects exist on stability when throats are inserted in compact spaces, see for instance \cite{Bena:2018fqc, Carta:2019rhx, Blumenhagen:2019qcg, Dudas:2019pls, Randall:2019ent}. 

\begin{figure}[h]
	\centering
	\includegraphics[width=0.6\linewidth]{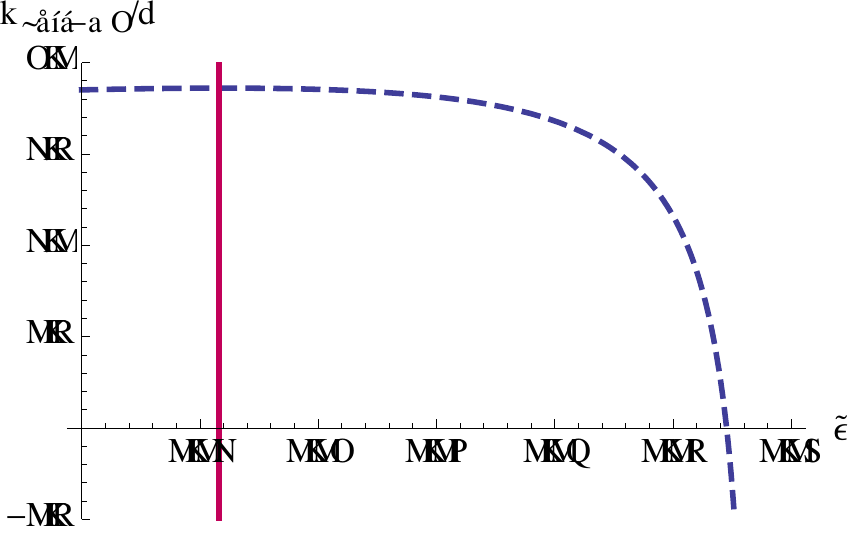}
	\caption{\small \emph{The behavior of the ratio $N_{\overline{\rm D2}} / G$ in units of $K \alpha^{-1}$ that determines brane-flux decay. The smaller the ratio the more certain one can be there is no decay. The region to the right of the vertical red line suffers from closed string tachyons.}} 
	\label{PLOTSTE}
\end{figure}

We have used anti-D2 branes and anti-D6 branes. Given the existence of smooth supersymmetric G2 holonomy throats with D2 charges that cap off in a finite $S^4$ \cite{Cvetic:2001ma}, we expect (in analogy with the KS throat) that also compact G2 manifolds can have such throats. The probe computation for brane-flux decay of anti-D2 branes has not yet been carried out\footnote{But results about the backreaction do exist at first order in the SUSY-breaking charge \cite{Giecold:2011gw}.} and is somewhat obscured by having to use the self-dual 3 forms on the IIA NS5 brane. We expect the computation to go along the lines of \cite{Bena:2000fz} and give a maximal value of $\frac{N_{\overline{\rm D2}}}{N_{\text{flux}}}$ of a few percent, just like for anti-D3 branes (\ref{anti-D3}) and anti-M2 branes (\ref{anti-M2}).  
This will then lead to a new problem for our 3d de Sitter vacua. 
Since equation \eqref{ND2-warped} implies
\be\label{problem}
\frac{N_{\overline{\rm D2}}}{G} = 7 K \alpha^{-1}\left( \frac12 - \frac{1}{6 \sqrt 2} 
\frac{4 - 43 \te + 18 \te^2}{
	\sqrt{4 - 71  \te + 18 \te^2} 
	(1 - \te) 
}  
\right) \, , 
\ee 
we will have to increase $\te$ such that the value of $N_{\overline{\rm D2}} / G$ 
drops and the system is safe from such decays (in our discussion until now we always assumed $\te\ll1$). 
However, 
we know that increasing $\te$ may lead to tachyons. 
Indeed, 
from \eqref{TildeE} we have $\te = 2 \epsilon/C$ and combined with \eqref{MMdet} we find that an absence of tachyons requires  (up to order $\te^2$) 
\be
\label{NOTACHYONS} 
\epsilon < C \frac{H-C}{39C + 4H} < 1 \ \ \to \ \ \frac{\te}{2} < \frac{H-C}{39 C + 4 H} \, . 
\ee
From now on we work only up to linear order in $\te$. This will turn out consistent since we will verify we need $\te < 0.012$. 
Indeed, using the properties of our solution and in particular \eqref{y0DS} we have 
\be
H = \frac{9}{7} (1 - 2 \te) C +  {\cal O}(\te^2)  \, . 
\ee
Then the no-tachyon condition \eqref{NOTACHYONS} reads 
\be
\label{NoTach}
 \frac{4 - 36 \te}{309 - 72 \te} - \te > 0 \, . 
\ee
One can see that the condition \eqref{NoTach} gives small values for $\te$ which essentially lead to large values of $N_{\overline{\rm D2}} / G$ from \eqref{problem}. 
This tension is depicted in figure \eqref{PLOTSTE}. 

To conclude, 
let us translate our findings in terms of $F_4$ flux quanta $G$. 
We have here a rather striking interplay between the parameters and the instabilities they relate to: 
trying to avoid brane-flux decay we have to increase $G$ which in turn works against a tachyon-free vacuum. 
Instead a small $G$ leads to tachyon-free vacua but then such vacua are jeopardized by brane-flux decay. 
So it seems that in our tachyon-free setup the fraction $N_{\overline{\rm D2}}/G$ is order one, 
and far well above in case there is warping. 
This however could be due to our isotropic way of stabilizing fluxes. 
We have not considered an alternative stabilization, 
but one would have to study different internal spaces that would naturally point towards a non-isotropic setup. 
It would be interesting to push this further and try to check whether this tension can somehow be relaxed.

Note that anti-D6 branes, wrapping 4d cycles will probably extend in the whole bulk. But it seems that the arguments of \cite{Blaback:2019ucp} leading to (\ref{anti-D6}) are model independent, although the actual bound might be tighter than (\ref{anti-D6}) due to backreaction. We have shown that in our model we could obey the bound (\ref{anti-D6}), but not parametrically.

\section{Outlook} 
We have shown that 3d flux compactifications of massive IIA supergravity with fluxes, O2/O6 and anti-D2/anti-D6 sources are an interesting environment for classical dS model building for multiple reasons:
\begin{itemize}
\item The standard tachyons (or better: reasons for tachyon) in classical dS vacua \cite{Danielsson:2012et, Junghans:2016uvg, Junghans:2016abx} seem absent for generic models. However flux quantization in the simplest toroidal model does lead to tachyons. As we argued throat models in which the anti-D2 branes have warped down tensions should be safe from this in case the throat can be long enough. The existence of such throats is inferred from the actual construction for non-compact G2 spaces \cite{Cvetic:2001ma}. 
\item The internal manifold can be Ricci flat and our understanding of the moduli problem is better for such manifolds.
\item If warping can be present one can achieve ``border-line'' numbers for the string coupling (order 0.1) and for the volume (order $10^7$ in string units)\footnote{So the radii of the separate circles are order 10.}.   
\end{itemize}

Note that anti-D6 uplifts have been found to be useful in 4d compactifications as well. For instance \cite{Cribiori:2019bfx, Cribiori:2019drf} claims they lead to a dS landscape but the construction involves a mixture of racetrack potentials and classical fluxes and it is unclear whether it is truly top down. On the other hand, reference \cite{Kallosh:2018nrk} added anti-D6 branes to the classical vacua of \cite{Danielsson:2011au} and found the tachyons were absent. But the problem of large coupling and small volume persists. What we suggest in this paper is that 3d compactifications allow classical solutions with better numbers and with flat internal spaces. Even more, if we use models with local throats the flux quantization problems we encountered seem alleviated. 

If the no-dS conjecture is correct the problem with this scenario for constructing dS solutions can come from multiple directions, 
which could actually work against each other. Maybe the issues we noticed with flux quantization bringing one away from the meta-stable minimum persist. 
Although as we argued, at the same level of precision of the uplift procedure in KKLT one could argue that sufficiently large throats can do the job. We gave an explicit example around equation (\ref{ND2-warped}) with very mild warping where already a meta-stable minimum was reached. A different problem with these vacua that could enforce the no-dS Swampland conjecture is the open string stability. We argued that the conditions for open string stability would be of the form (\ref{openstringstability}). We have explicitly verified that the stability of the anti-D6 was satisfied in case we can trust existing probe results \cite{Blaback:2019ucp} but that the stability of the anti-D2 is a worry due to equation (\ref{problem}), 
although no concrete brane-flux decay computation is carried out in lack of a concrete model without closed string tachyons after flux quantization. 
At this point it makes sense to assume that the no-dS conjecture would actually be enforced from the open-string sector once the closed-string sector seems stable. So dS model building continues to share many analogies with the ``whack-a-mole'' game.

\section*{Acknowledgements} 
We would like to thank Iosif Bena for a useful discussion. 
The work of FF and TVR is supported by the KU Leuven C1 grant ZKD1118C16/16/005.

\appendix

\small

\providecommand{\href}[2]{#2}\begingroup\raggedright\endgroup

\end{document}